\documentclass{article}
\usepackage{ijcai24}

\usepackage{times}
\usepackage{soul}
\usepackage{url}
\usepackage[hidelinks]{hyperref}
\usepackage[utf8]{inputenc}
\usepackage[small]{caption}
\usepackage{graphicx}
\usepackage{booktabs}
\usepackage{algorithm}
\usepackage{algorithmic}
\usepackage[switch]{lineno}

\usepackage{amsmath}
\usepackage{amsthm}

\usepackage{caption}
\usepackage{subcaption}

\usepackage[title]{appendix}

\providecommand{\citet}[1]
{
  \citeauthor{#1} (\citeyear{#1})
}

\title{Enabling Sustainable Freight Forwarding Network via Collaborative Games}

\author{
Pang-Jin Tan$^1$
\and
Shih-Fen Cheng$^1$\and
Richard Chen$^2$\\
\affiliations
$^1$Singapore Management University\\
$^2$Coupang\\
\emails
\{pangjin.tan.2021, sfcheng\}@smu.edu.sg,
richen2@coupang.com
}

\begin{document}

\maketitle

\begin{abstract}
Freight forwarding plays a crucial role in facilitating global trade and logistics. However, as the freight forwarding market is extremely fragmented, freight forwarders often face the issue of not being able to fill the available shipping capacity. This recurrent issue motivates the creation of various freight forwarding networks that aim at exchanging capacities and demands so that the resource utilization of individual freight forwarders can be maximized. In this paper, we focus on how to design such a collaborative network based on collaborative game theory, with the Shapley value representing a fair scheme for profit sharing. Noting that the exact computation of Shapley values is intractable for large-scale real-world scenarios, we incorporate the observation that collaboration among two forwarders is only possible if their service routes and demands overlap. This leads to a new class of collaborative games called the Locally Collaborative Games (LCGs), where agents can only collaborate with their neighbors. We propose an efficient approach to compute Shapley values for LCGs, and numerically demonstrate that our approach significantly outperforms the state-of-the-art approach for a wide variety of network structures.
\end{abstract}

\section{Introduction}

Freight forwarding plays a crucial role in facilitating global trade and logistics. In particular, freight forwarders often act as intermediaries between shippers and carriers, offering one-stop services that would ensure smooth and efficient transportation of goods. However, as the freight forwarding market is extremely fragmented~\cite{ForwardingIndustryReport,FreightOS}, freight forwarders often face the issue of not being able to fully fill the available shipping capacity. This recurrent issue motivates the creation of various freight forwarding networks that aim at exchanging capacities and demands so that the resource utilization of individual freight forwarders can be maximized. In this paper, we focus on how to design such a collaborative network, emphasizing deriving a fair profit-sharing scheme that is firmly rooted in the collaborative game theory. 

The collaboration between freight forwarders can be modeled as a cooperative game, with a strong network structure that is not present in classical cooperative games. More specifically, in the classical cooperative game theory, we assume that agents can collaborate with each other without restriction. However, there are realistic situations where agents can only collaborate with some other agents but not necessarily with all of the rest. This has given rise to graph-restricted games. \citet{myerson1977graphs} further proposed a solution concept that allows for a value function defined only for connected coalitions. \citet{amer2004connectivity} proposed an alternative game called connectivity game which also addresses the possibility of agents not fully collaborating. \citet{skibski2014algorithms} developed an approach that combines a fast algorithm for enumerating induced subgraphs along with a counting approach to efficiently compute the Myerson value of an arbitrary graph-restricted game.

Our work in this paper is closely related to \citet{skibski2014algorithms} in that we also rely on an underlying graph structure and a counting approach to compute Shapley value, but we do it for a certain class of graph-restricted games that are suitable for the construction of the freight forwarding network. Our contributions can be summarised as follows:
\begin{itemize}
    \item We propose the Locally Collaborative Game (LCG) that can be used to represent a general class of collaborative games that comes with a strong network property.

    \item We exploit the properties of LCG to derive an efficient algorithm to compute the Shapley value as compared to the approach proposed by \citet{skibski2014algorithms}.
    
    \item Finally, we apply the algorithm to the freight forwarder network. We define the Freight Forwarders Collaboration Game (FFCG), which is an instance of LCG, and show that our algorithm outperforms the baseline method in a general setting as well as in two practical scenarios, one with a highly fragmented market and one where there exists a few dominant market players. 
\end{itemize}

\section{Background}

We first establish some graph theoretic notations. A graph $G=(V,E)$ is defined by $V$, the set of vertices (or nodes), and $E$, the set of edges where each edge $e \in E$ is connected by two vertices $u, v \in V$. An induced subgraph $G'=(V', E')$ is formed by selecting a subset of vertices $V' \subseteq V$ and including all edges from $E$ such that their endpoints are vertices in $V'$. An induced subgraph $G'=(V', E')$ is connected if there exists a path from any vertex $u \in V'$ to any other vertex $v' \in V'$.

We next review fundamental concepts from cooperative game theory. A cooperative game $(N,v)$ is characterized by $N$, the set of agents (or players), and $v$, the characteristic function describing the value of a given coalition. Formally, $v:2^N \rightarrow \mathbf{R}$ is a function that assigns a real value to all possible subsets of $N$. The characteristic function $v$ tells us the payoff for a coalition of agents when they collaborate. A game is super-additive if for any two disjoint coalitions $S$ and $T$, $v(S \cup T) \geq v(S) + v(T)$. In other words, the value generated by two groups collaborating is at least as much as the sum of the values generated by each group acting independently. This property ensures that agents are incentivized to form larger coalitions, leading to the formation of the grand coalition where all agents collaborate.

Suppose we have a grand coalition formed, then we have a total value of this collaboration given by the characteristic function. An important question arises as to how we allocate this total value to individual agents in a fair manner. One such allocation mechanism is the Shapley value. It is the unique allocation scheme that satisfies the properties of efficiency, symmetry, null player, and additivity. Specifically, the Shapley value for agent $i$ is
\begin{align}
    Sh_i(v) = \frac{1}{|N|!} \sum_{\pi \in P(N)} v\bigl( S_{\pi}(i) \cup \{i\} \bigr) - v\bigl( S_{\pi}(i) \bigr) \;, \label{shapley}
\end{align}
where $N$ is the set of agents, $P(N)$ is the set of all permutations of $N$ and $S_{\pi}(i)$ is the set of agents which are predecessors of $i$ in the specific permutation $\pi$.

In other words, we can think of Shapley value for agent $i$ to be the average marginal contribution that agent $i$ brings to its predecessors across all possible permutations for which agents can join to form the grand coalition.

\section{Locally Collaborative Game}

In this section, we define the Locally Collaborative Game (LCG) and in the next section we propose an efficient approach to calculate its Shapley value we call Fast Shapley for Locally Collaborative Game (FS-LCG). 

Let $N$ be a set of agents, and $G=(N, E)$ be the graph depicting the collaboration structure between the agents in $N$. In other words, there is an edge between two agents if and only if the agents can collaborate. We call $(N,v)$ a Locally Collaborative Game (LCG) if the characteristic function satisfies the following \textit{locally collaborative property}:
\begin{multline}
    v(\{x\} \cup X) - v(X) = v(\{x\} \cup X') - v(X'),\\
    \forall x\in N, \forall X \subseteq N \setminus \{x\}, 
\end{multline}
where $X'\subseteq X$ and $X'$ is the neighbors set of $x$. In other words, the marginal contribution of $x$ to a set of agents $X$ is only affected by $x$'s neighbors in $X$.

\begin{figure}[tb]
  \centering
  \includegraphics[width=2in]{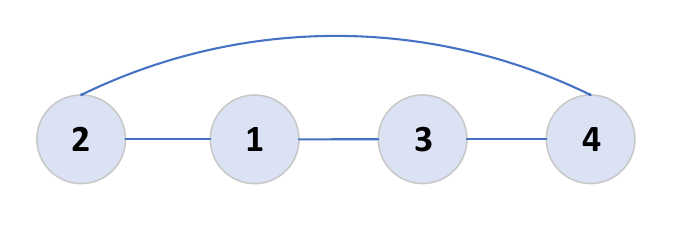}
  \caption{An LCG example.}
  \label{lcg_eg}
\end{figure}

As an example, suppose we have an LCG given by the collaboration graph in Figure \ref{lcg_eg}, then the marginal contribution of agent $1$ to the coalition $\{2,4\}$ is the marginal contribution of $1$ to $\{2\}$ since agent $4$ is not connected to $1$. 

Similarly, the marginal contribution of agent $1$ to coalition $\{2,3,4\}$ is the marginal contribution of $1$ to coalition $\{2,3\}$. Mathematically, we have $v(\{1,2,3,4\})-v(\{2,3,4\}) = v(\{1,2,3\}) - v(\{2,3\}) = v(\{1,2,3\}) - v(\{2\}) - v(\{3\})$. Note that the last equality also makes use of the locally collaborative property, where $v(\{2,3\})-v(\{3\})=v(\{2\})-v(\emptyset)$. More generally, the locally collaborative property also implies that $\forall X, Y \subseteq N$, $X$ and $Y$ are not connected, we have $v(X \cup Y) = v(X) + v(Y)$. In other words, when two disjoint coalitions form a bigger coalition, there is no marginal increase in the total payoff.

\section{Shapley Value for the LCG}

In this section, we propose an algorithm to compute the Shapley value for the LCG. Recall the expression of Shapley value as we stated in equation \ref{shapley}. In this approach, we consider all permutations in which agents join to form a grand coalition. We consider for each permutation, the marginal contribution of agent $i$, and compute the average across all permutations. Given a particular permutation $\pi$ and an agent $i$, there are two cases to consider. In the first case, suppose there is no agent in $S_{\pi}(i)$ which is a neighbor of $i$, then the marginal contribution of agent $i$ to $S_{\pi}(i)$ is simply $v({i})$, and the number of such permutations is
\begin{align}
    A(i)=\sum^{|N|-n_i-1}_{k=0} {|N|-n_i-1 \choose k} k! \bigl( |N|-k-1 \bigr)! ,
\end{align}
where $n_i$ is the number of agents connected to $i$. The first factor in each term accounts for the number of ways we can choose non-neighbors of $i$, the second factor accounts for the number of ways we can permutate these non-neighbors in front of $i$, and the last factor accounts for the number of ways we can permutate the rest of the agents.

In the second case, suppose there is at least one neighbor of $i$ in $S_{\pi}(i)$. Let $S'_{\pi}(i) \subseteq S_{\pi}(i)$ be the neighbor set of $i$. Then by locally collaborative property, the marginal contribution of $i$ to $S_{\pi}(i)$ is:
\[
v\bigl(S_{\pi}(i) \cup {i}\bigr) - v\bigl(S_{\pi}(i)\bigr) = v\bigl(S'_{\pi}(i) \cup {i}\bigr) - v\bigl(S'_{\pi}(i)\bigr).
\]
The number of such permutations is:
\begin{multline}
    B(i)=\sum^{|N|-n_i-1}_{k=0} {|N|-n_i-1 \choose k} \bigl( k+|S'_{\pi}(i)| \bigr)!\\
    \bigl(|N|-k-1-|S'_{\pi}(i)| \bigr)! \;,
\end{multline}
where the first factor accounts for the number of ways we can choose non-neighbors of $i$, the second factor accounts for the number of ways we can permutate these non-neighbors along with the neighbors of $i$, and the last factor accounts for the number of ways we can permutate the rest of the agents. 

We can now write a general expression for Shapley value for LCG as follows:
    \begin{multline*}
         Sh_i^{LCG}(v)^ = \frac{1}{|N|!} \sum_{\pi \in P(N)} \mathbf{1}(S_{\pi}(i) \cap N(i)) v(\{i\}) A(i) \\
         + (1 - \mathbf{1}(S_{\pi}(i) \cap N(i)) )(v(S^i_{\pi}(i) \cup \{i\}) - v(S^i_{\pi}(i)))B(i)   
    \end{multline*}
    where
        \[
        \mathbf{1}(S) = \begin{cases}
        1 & \text{if } S = \emptyset, \\
        0 & \text{otherwise.}
        \end{cases}
        \]

\begin{figure}[tb]
  \centering
  \includegraphics[width=3.375in]{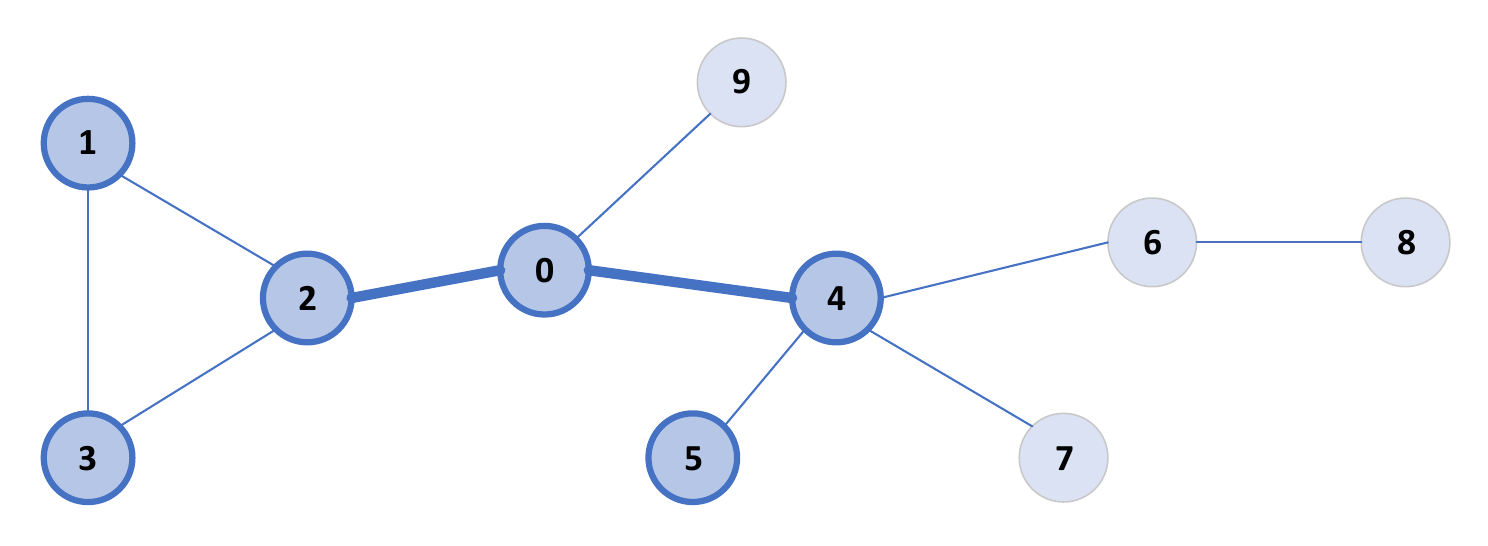}
  \caption{10-agent LCG example.}
  \label{lcg_10_ff}
\end{figure}

As an illustration, consider a set of $10$ agents depicted by the collaboration graph in Figure \ref{lcg_10_ff}. Given the permutation 1,2,3,4,5,0,6,7,8,9, the marginal contribution of $0$ to its predecessors is $v(\{1,2,3,4,5,0\})-v(\{1,2,3,4,5\})$ which simplifies to $v(\{2,4,0\}) - v(\{2,4\})$ since $2,4$ are the only neighbors of agent $0$ which appear in front of it. To compute how many permutations will give rise to the marginal contribution of $v(\{2,4,0\}) - v(\{2,4\})$, we first note that there are $6$ non-neighbors of agent $0$. Then we choose either $0,1,2,3,4,5$ or $6$ of them along with agents $2,4$ (which we must choose), and we permutate them in front of agent $0$. Finally, we permutate the rest of the agents behind agent $0$. The total number of such permutations is $\sum^{6}_{k=0} {6 \choose k}(k+2)!(9-k-2)!$.

There are two key ideas that we have discussed so far. First, only the neighbors of an agent matter when we consider the marginal contribution of that agent. Second, there are many permutations that would give rise to the same marginal contribution, hence we only need a way to count them. With these ideas, we describe our algorithm Fast Shapley for Locally Collaborative Game (FS-LCG) in Algorithm \ref{alg:FS-LCG}. For each agent $i$, for each subgraph $G'$ induced by $i$'s neighbors set $N_i$, compute marginal contribution $v(G'\cup {i})-v({G'})$ and its weight given by $\sum^{|N|-n_i-1}_{k=0} {|N|-n_i-1 \choose k}(k+|G'|)!(|N|-k-1-|G'|)!$. The Shapley value for agent $i$ is then the weighted average of the marginal contribution of $i$ to all the coalitions formed by agents of its neighbors.

\begin{algorithm}[tb]
    \caption{Fast Shapley for LCG (FS-LCG)}
    \label{alg:FS-LCG}
    \textbf{Input}: $G=(N,E),v$\\
    \textbf{Output}: $Sh_i(v)$
    \begin{algorithmic}[1] 
        \FORALL{agent $i \in N$}
            \STATE $N_i \leftarrow$ neighbors of $i$
            \STATE $n_i \leftarrow$ $|N_i|$
            \STATE $\mathcal{P}(N_i) \leftarrow$ set of induced subgraphs with nodes in $N_i$
            \FORALL{induced subgraph $G' \in \mathcal{P}(N_i) $}
                \STATE $mc$ $\leftarrow v(G'\cup \{i\}) - v(G')$
                \STATE $w$ $\leftarrow \sum^{|N|-n_i-1}_{k=0} {|N|-n_i-1 \choose k}(k+|G'|)!(|N|-k-1-|G'|)!$
                \STATE $Sh_i(v) \leftarrow Sh_i(v) + \frac{1}{|N|!} \times w \times mc$
            \ENDFOR
        \ENDFOR
    \end{algorithmic}
\end{algorithm}

\section{The Freight Forwarder Collaboration Game (FFCG)}

Freight forwarders are essential intermediaries in today's complex global supply chain as they are the critical bridges between shippers and carriers. They first procure transportation capacities from carriers, such as airlines or shipping lines, and then resell them to their clients (i.e., the shippers), often bundled with additional services like customs brokerage. Forwarders can negotiate preferential rates with carriers because they purchase large quantities of capacity. The rates are typically much lower than the spot rates available to shippers. Therefore, for shippers, working with a freight forwarder instead of a carrier can simplify the shipping process and save them money. However, freight forwarders have their own challenges in managing capacities. It's difficult for them to accurately forecast capacity needs due to uncertain demand from shippers. Procuring too much capacity reduces profits, while insufficient capacity leads to lost sales. 

To address the systemic inefficiencies in the freight forwarding industry, we propose a digital freight marketplace for forwarders. This platform helps participating forwarders reduce their operating costs by facilitating collaboration among them. Forwarders first submit their available capacities and transport requests to the platform. The platform will then optimally reallocate requests to available capacities across all the forwarders. This will result in a lower total shipping cost for all requests compared to the cost incurred when each forwarder only uses its own capacities. This process of assigning requests to capacities can be formulated as an integer linear program. To solve large-scale scenarios, we use a two-step approach that combines a greedy algorithm with an exact fine-tuning step. Finally, after we have found the optimal overall shipping cost for all requests, we need an equitable mechanism to allocate the total cost to each forwarder. We model this problem as a cooperative game and compute the Shapley value allocation. The game turns out to be an instance of LCG and we will use the algorithm FS-LCG for computing the Shapley value.

\subsection{Motivation}

In this subsection, we discuss a specific form of collaboration for forwarders and the potential benefits they can derive. As mentioned earlier, freight forwarders are intermediaries between shippers and carriers. They negotiate contracts with carriers to secure capacity at discounted rates, typically twice a year for the winter and summer seasons. Forwarders analyze historical data to predict demand for different service port pairs. An example of a service port pair is USLAX-HKHKG representing service from Los Angeles to Hong Kong. When the freight season begins, forwarders use the procured capacity to fulfill shippers' transport requests.

In this paper, we focus on Less-than-Container-Load (LCL) ocean freight shipments, where shippers need only a partial space of a container instead of a full container. To better utilize their contracted capacity, forwarders consolidate multiple LCL shipments of different sizes from multiple shippers into a single container. The more efficient the consolidation strategy, the more profitable the forwarder is. The consolidation problem can be modeled as a one-dimensional bin-packing problem.

We now look at a concrete example to illustrate the points above. Consider Forwarder A (see Table \ref{table_1_2}), which has a demand of four requests and a supply of three containers. Assuming a 20ft container with a capacity of 30 cubic meters (cbm), requests 1 and 2 can be consolidated into one container, but requests 3 and 4 must be shipped in separate containers. This results in a total incurred cost of \$3000.

\begin{table}[tb]
\begin{center}
\resizebox{0.4\textwidth}{!}{
\begin{tabular}{|c||c||c|}
\multicolumn{3}{c}{Demand}\\
\hline
Request & Volume (cbm) & Service Required\\
\hline
1 & 14 & USLAX-CNSHA\\
2 & 12 & USLAX-CNSHA\\
3 & 10 & USLAX-CNSHA\\
4 & 15 & DEHAM-SGSIN\\
\hline
\multicolumn{3}{c}{Supply}\\
\hline
Service & Cost per container & Supply\\
\hline
USLAX-CNSHA & \$900 & 2\\
DEHAM-SGSIN & \$1200 & 1\\
\hline
\end{tabular}}
\caption{Forwarder A's demand \& supply.}
\label{table_1_2}
\end{center}
\end{table}

\begin{table}[tb]
\begin{center}
\resizebox{0.4\textwidth}{!}{
\begin{tabular}{|c||c||c|}
\multicolumn{3}{c}{Demand}\\
\hline
Request & Volume (cbm) & Service Required\\
\hline
1 & 6 & USLAX-CNSHA\\
2 & 6 & USLAX-CNSHA\\
3 & 6 & USLAX-CNSHA\\
4 & 6 & USLAX-CNSHA\\
5 & 15 & DEHAM-SGSIN\\
\hline
\multicolumn{3}{c}{Supply}\\
\hline
Service & Cost per container & Supply\\
\hline
USLAX-CNSHA & \$1000 & 2\\
DEHAM-SGSIN & \$1100 & 1\\
\hline
\end{tabular}}
\end{center}
\caption{Forwarder B's demand \& supply.}
\label{table_3_4}
\end{table}

In practice, forwarders often find it difficult to manage demand uncertainty resulting in over-supply of some services but under-supply of others. However, collaboration between forwarders can help alleviate these issues. Here is an illustration. Table \ref{table_3_4} shows Forwarder B's demand and supply. Without collaboration, Forwarder B would incur a total cost of \$2100, and the combined cost for Forwarders A and B would be \$5100. Now we consider what happens if Forwarders A and B collaborate by sharing capacities. Note that the first three requests from Forwarder A and the first four requests from Forwarder B are all shipped from USLAX to CNSHA. We can minimize costs by assigning requests 1 and 3 from Forwarder A and request 1 from Forwarder B to the first container that belongs to Forwarder A. We then assign request 2 from Forwarder A and requests 2, 3, and 4 from Forwarder B to the second container of Forwarder A. The total shipping cost for USLAX-CNSHA is \$1800. Likewise, both DEHAM-SGSIN requests can be consolidated into a container procured by Forwarder B for a cost of \$1100. The combined incurred cost is \$2900, which is significantly lower than the case without collaboration.

\subsection{Optimal Assignment of Requests to Boxes}

Generalizing from the example described in the earlier subsection, we now formulate an integer linear program called Freight Forwarders Collaboration Problem (FFCP), which describes how forwarders can collaborate by sharing capacities to achieve a lower total cost.

\begin{flushleft}
\textbf{Index Sets}
\end{flushleft}
$R$ is the set of requests, indexed by $r$. $S$ is the set of services, indexed by $s$. $R_s$ is the set of requests that can be assigned to $s$. $S_r$ is the set of services that is feasible for $r$.

\begin{flushleft}
\textbf{Parameters}
\end{flushleft}
$c_s$ is the cost per container on service $s$. $n_s$ is the number of containers available on service $s$. $v_r$ is the volume of request $r$. $v^{max}$ is the max volume of a box.

\begin{flushleft}
\textbf{Decision Variables}
\end{flushleft}
$x^i_{r,s}$ is $1$ if request $r$ is shipped on service $s$ box $i$, $0$ otherwise. $y^i_s$ is $1$ if service $s$ box $i$ is used, $0$ otherwise.

\begin{flushleft}
\textbf{Model FFCP(R,S)}
\end{flushleft}
\begin{align}
    & \min \sum_{s \in S} \sum_{i=1}^{n_s} c_s y^i_s \\
\text{s.t.}& \nonumber \\
    & \sum_{s \in S_r} \sum_{i=1}^{n_s} x^i_{r,s} = 1, \forall r \in R,   \label{m2_c1} \\
    & \sum_{r \in R_s}  v_r x^i_{r,s} \le v^{max} y^i_s, \forall s \in S, i=1,...n_s, \label{m2_c2} \\
    & x^i_{r,s}, y^i_s \in \{0,1\}.  \label{m2_c3}
\end{align} 

Given a set of forwarders $F$, let $R_f$ be the set of requests to be handled by forwarder $f\in F$ and let $S_f$ be the set of services procured by forwarder $f\in F$. Let $R=\cup_f R_f$ be the set of requests to be handled by all forwarders in $F$, and $S=\cup_f S_f$ be the set of services procured by all forwarders $F$. Then the optimization problem FFCP(R,S) parameterized by the set of requests $R$ and set of services $S$ is given above. The objective is to minimize the total cost of shipping all requests. Constraint \eqref{m2_c1} ensures that each request is assigned to exactly one of the containers. Constraint \eqref{m2_c2} ensures that the total volume of requests fitted in a container does not exceed maximum capacity. Constraint \eqref{m2_c3} ensures assignment variables are binary. Finally, we define $\phi(R,S)$ to be the optimal objective value obtained by solving FFCP(R,S).

\subsection{Equitable Cost Allocation for Forwarders}

When multiple forwarders collaborate to share capacities, their total cost is reduced. An important question that arises is how the total cost should be allocated to the forwarders. Clearly, different forwarders, by virtue of their procured capacities, bring different values to a coalition. We need a fair allocation scheme where two forwarders are allocated the same cost if they provide the same savings when added to a coalition of forwarders. This suggests that we use the Shapley value as our allocation method. 

Towards computing the Shapley value, our next step is to formulate the Freight Forwarders' Collaboration Game (FFCG) for the forwarders, $\langle F, v \rangle$, where the characteristic function $v:2^F \rightarrow \mathbf{R}$ assigns a real value to every subset of $F$ such that $v(\tilde{F)}=\phi(\tilde{R},\tilde{S})$, where $\tilde{F} \subseteq F$, $\tilde{R}$ is the set of requests handled by forwarders in $\tilde{F}$ and $\tilde{S}$ is the corresponding services procured by forwarders in $\tilde{F}$. In other words, the value of coalition arising from collaborating forwarders is the minimum cost of shipping all the requests of the collaborating forwarders given the services they have procured.

The game $\langle F, v \rangle$ is super-additive since for any two forwarders, each of them is never worse off by collaborating with one another. Each of them potentially has access to a cheaper capacity to ship its requests. At worst, it simply uses its own capacities. The superadditivity of the game thus implies that the grand coalition will form, i.e., all forwarders are incentivized to collaborate in a single coalition. Furthermore, we can assume that the grand coalition of forwarders is stable since in reality, it is not easy for forwarders to break away from the grand coalition to form sub-coalitions as it requires the exchange of information between the forwarders.

\section{Solution Approach for the FFCG}

To compute the Shapley value for forwarder $i$ in FFCG, there are two critical parts to it. First, the computation of characteristic function $v$ involves solving FFCP which is a variant of the bin-packing problem, hence NP-Hard. In this work, we follow \citet{10371994} to exploit a decomposability property and apply a greedy approach to speed up computation. As an illustration, the situation described in \ref{table_1_2} and \ref{table_3_4} can be recast and depicted in \ref{graph2}. Second, we exploit the fact that FFCG is an instance of LCG to speed up the computation of the Shapley value.

\subsection{Computing Shapley Value Efficiently}

To compute the Shapley value for FFCG, we first note that FFCG is an instance of LCG, hence we can use our earlier proposed approach of FS-LCG. To see why FFCG is an instance of LCG, note that for any two forwarders to collaborate, they must have some services that serve the same port pairs. Given a coalition of some forwarders $X$, suppose a new forwarder $x$ joins, those forwarders in $X$ which have no common port pairs with $x$ (call them $X''$) will not play a role in affecting the marginal contribution of $x$ to $X$, because the transport requests in these forwarders $X''$ cannot be handled by $x$ and vice versa. In other words, there is no synergy between $x$ and $X''$. Since FFCG is an instance of LCG, we utilize FS-LCG. To compute the Shapley value for a given forwarder, we find all subgraphs induced by their neighbors, and compute their marginal contributions to a given subgraph, along with the weighting factor.

\begin{figure}[tb]
  \centering
  \includegraphics[width=3.375in]{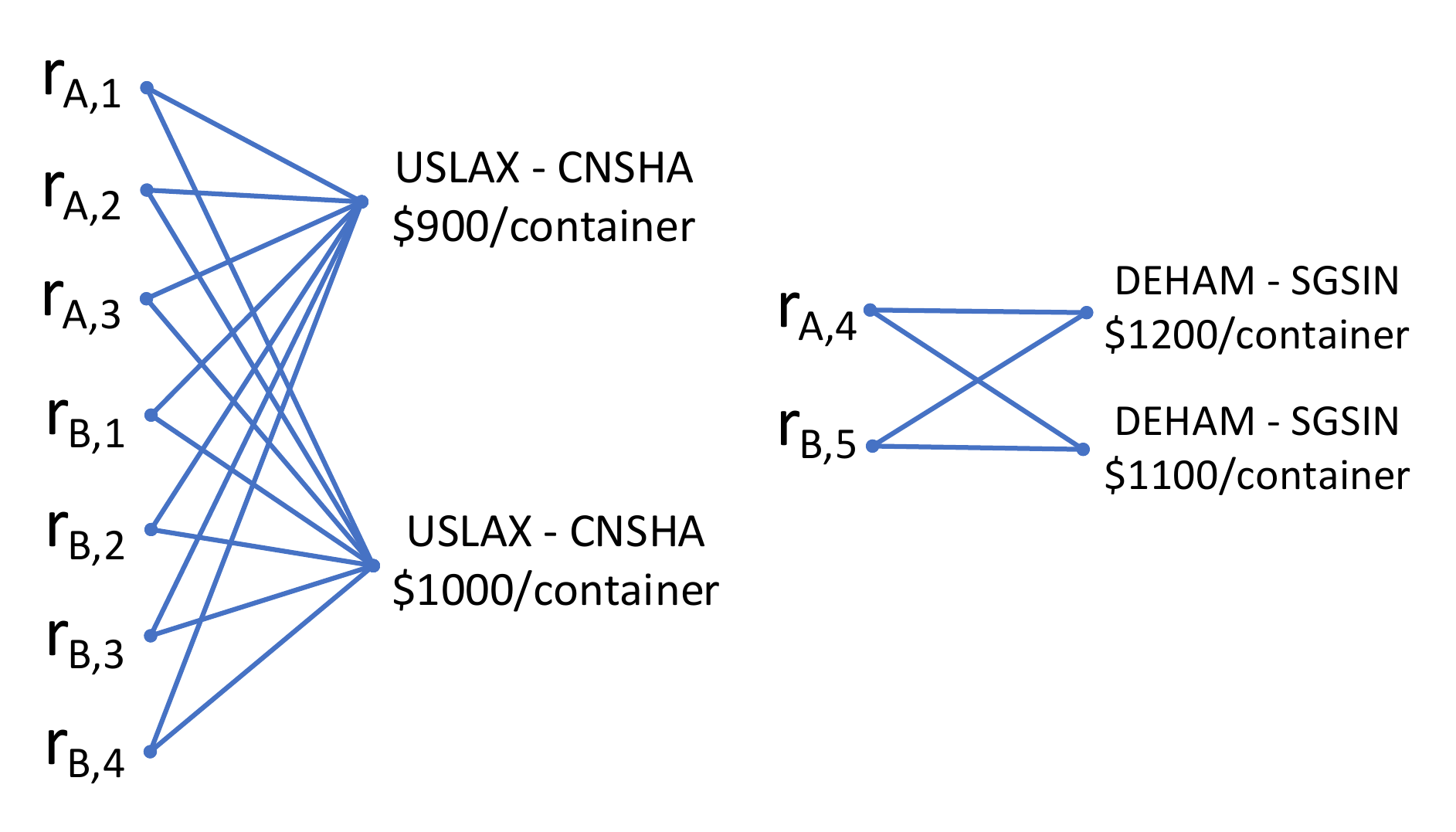}
  \caption{Request-service assignments decomposed into non-overlapping sub-problems.}
  \label{graph2}
\end{figure}

\begin{figure}[tb]
  \centering
  \includegraphics[width=3.375in]{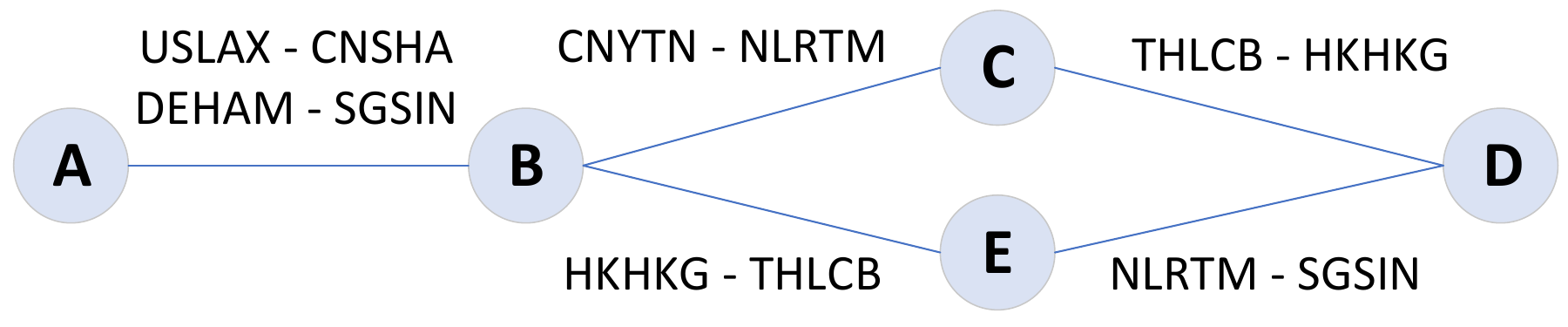}
  \caption{An FFCG example.}
  \label{ffcg_eg}
\end{figure}

Here's an illustration. Figure \ref{ffcg_eg} shows five forwarders where each of them has a set of transport requests for the services that they have procured. The port pair on an edge between two nodes refers to the service that both nodes (forwarders) have to serve. The edge between A and B implies that Forwarders A and B have requests for port pairs USLAX-CNSHA and DEHAM-SGSIN. Transport requests, service capacity, and cost are given in the FFCP described in the earlier section as depicted in Figure \ref{graph2}. Likewise, the edge between B and C represents the fact that both Forwarder B and C have requests for port pair CNYTN-NLRTM. Similar interpretations apply to other edges on the graph.

We now apply FS-LCG to compute the Shapley value for Forwarder A. Note that we only need to compute $v(\{A\})$ and $v(\{A,B\})$ since B is the only node connected to A. Based on the parameter data in the previous subsection, we have $v(\{A\})=3000$ and $v(\{A,B\})=2900$. Furthermore, coalition values generated by other subgraphs are not relevant.
\begin{itemize}
    \item First case: we consider permutations where none of the Forwarders in front of A is a neighbor. There are 3 nodes that are non-neighbors of A and we can choose either 0,1,2,3 of them and permutate them in front of A. We then permutate the rest after A. The total number of such permutations is ${3 \choose 0}0!4!+{3 \choose 1}1!3!+{3 \choose 2}2!2!+{3 \choose 3}3!1!=60$. The marginal contribution of agent A in all these cases is $v(\{A\})=3000$.
    
    \item  Second case: we consider permutations where there is at least a forwarder in front of A which is a neighbor. In this case, the only neighbor of A is B, hence we must place B in front of A. Furthermore, we can choose any of the non-neighbors of A to be in front of A. The total number of such permutations is ${3 \choose 0}1!3!+{3 \choose 1}2!2!+{3 \choose 2}3!1!+{3 \choose 3}4!0!=60$. The marginal contribution of agent A in all these cases is $v(\{A,B\})-v(\{B\})=-100$.
    
    \item Hence, Shapley value for Forwarder A is $\frac{1}{5!} \times (60 \times 3000 - 60 \times 100) = 1450$ .
\end{itemize}

\section{Numerical Experiments}

To analyze the applicability of our approach, we set up three different experimental scenarios. In the first experiment, the objective is to understand the performance of FS-LCG in a general setting. We have $100$ port pairs with the number of services per port pair set range between 1 and 5. We then randomly distribute these services uniformly across the forwarders. The cost of a box is drawn from $U(700,1300)$, the number of boxes for a service is drawn from $U(20,80)$, the number of transport requests per service is drawn from $U(1,5)$, and the volume of transport request is drawn from $U(1,29)$. We first generate profiles for $30$ forwarders. Subsequently, to create new instances, we repeatedly remove $5$ forwarders at one time.

In the second set of experiments, to reflect the real-world scenarios where the freight forwarder market is highly fragmented, we generate small-world graphs of various sizes and degrees. The average degrees range from 2 to 8 whereas the number of forwarders ranges from 10 to 50. To generate small-world graphs, we start with a regular graph based on the number of nodes and average degrees. We randomly select an edge with a probability of 0.2 and rewire one node to another node. For each edge, we randomly generate services. Other parameters are the same as those used in the first set of experiments.

In the third set of experiments, we generate graphs where the degrees follow a power-law distribution. This is to simulate the scenarios where there exists a few dominant key players with relatively higher market share. The number of nodes ranges from 5 to 40 with a power-law exponent of 2. Other parameters remain the same as used in the first set of experiments. 

We compare our approach against a baseline method developed by \citet{skibski2014algorithms}, which was developed to compute Shapley value for graph-restricted games in a general setting. More specifically they studied graph-restricted games but assumed that two agents could co-operate as long as there is a path between them. Furthermore, they did not consider a specific form of characteristic function but rather assumed that it could be calculated in constant time.

\subsection{Results}

\begin{figure}[tb]
  \centering
  \includegraphics[width=3.375in]{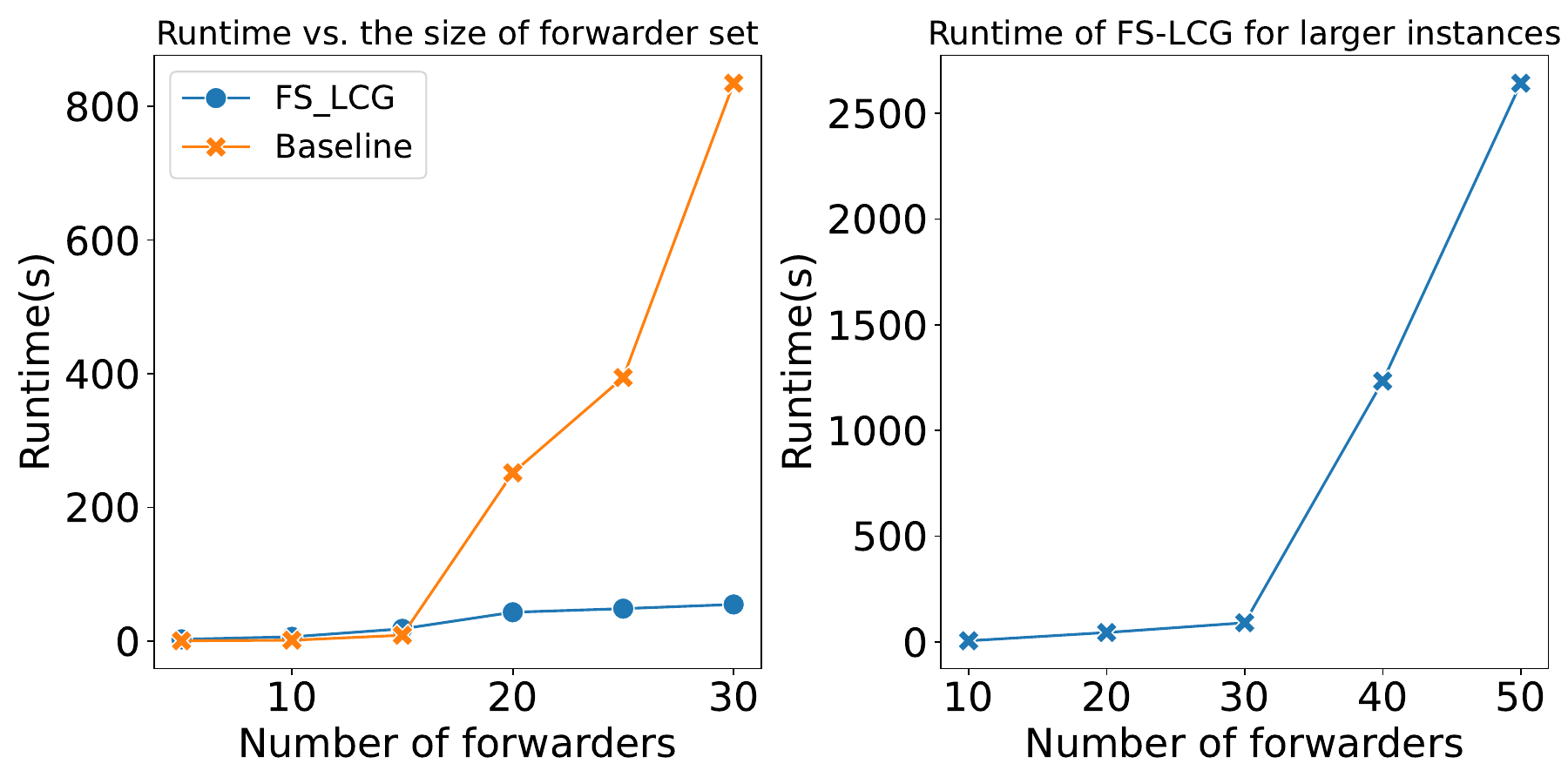}
  \caption{Runtime in a general setting.}
  \label{res1}
\end{figure}

In our first experiment, we compare the runtime between FS-LCG and the baseline method by \citet{skibski2014algorithms} as we vary the number of forwarders from $5$ to $30$. Our method is comparable in performance compared to the baseline when the number of forwarders is fewer than $15$, as shown in Figure \ref{res1}. Furthermore, when there are more than $15$ forwarders, FS-LCG outperforms the baseline significantly, leading to runtime reductions of around $82\%$ to $93\%$.

We then investigate the performance of FS-LCG as we increase the size of the forwarder set to a reasonably big set. Our method can still perform reasonably well within practical limits as shown in Figure \ref{res1}, where the runtime reaches around 43 mins for the case with 50 forwarders.

\begin{figure}[tb]
  \centering
  \includegraphics[width=3.375in]{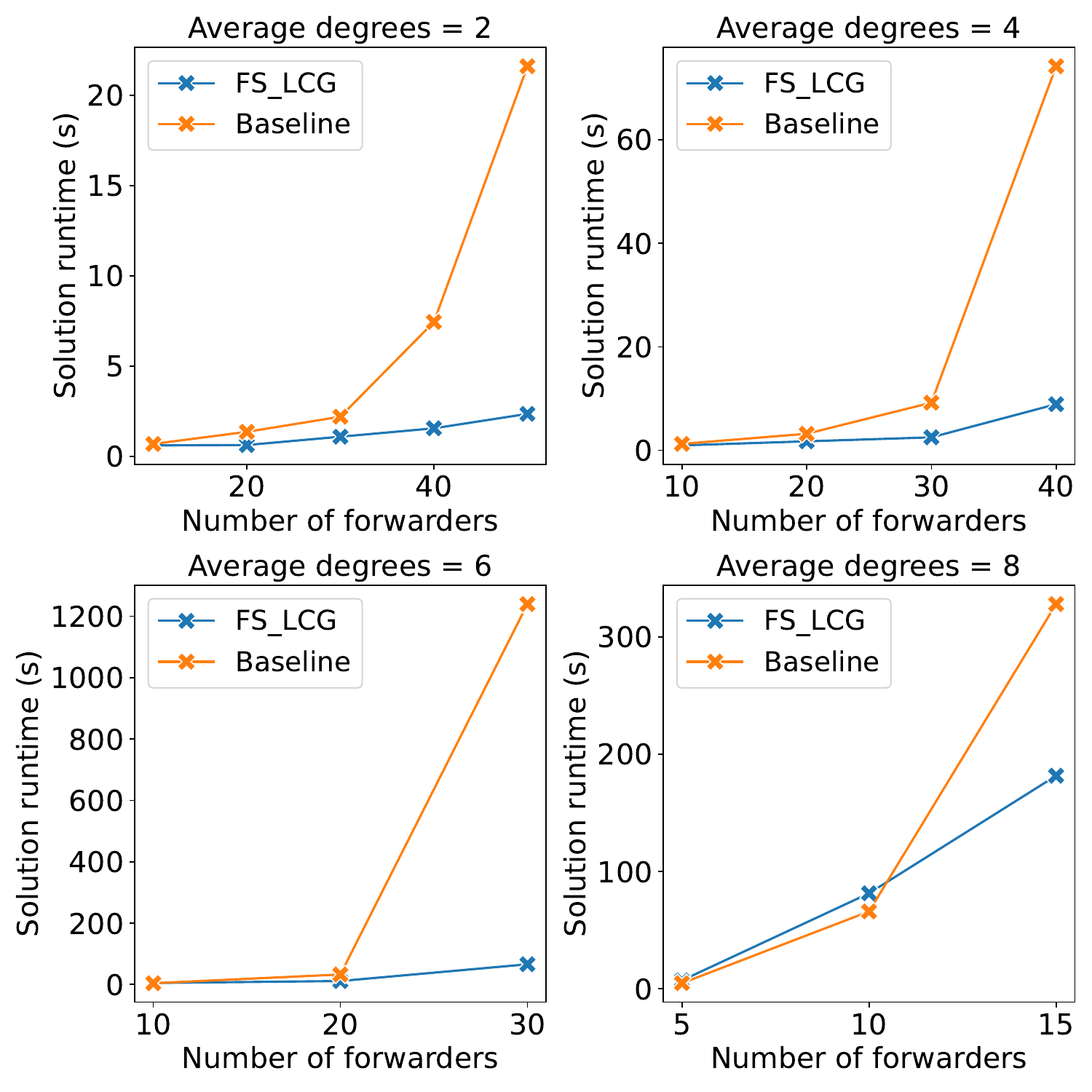}
  \caption{Runtime for small-world graphs.}
  \label{res2}
\end{figure}

In the second experiment, we observe that the higher the number of forwarders, the better the performance of FS-LCG as compared to the baseline method, (see Figure \ref{res2}) with a maximal saving of 94\% and a minimal saving of 11\%. This improvement is consistent across the scenarios with different average degrees. Furthermore, we also note that for the same number of forwarders, the improvement in runtime is more significant as the average degrees of the graphs increase. 

\begin{figure}[tb]
  \centering
  \includegraphics[width=3.375in]{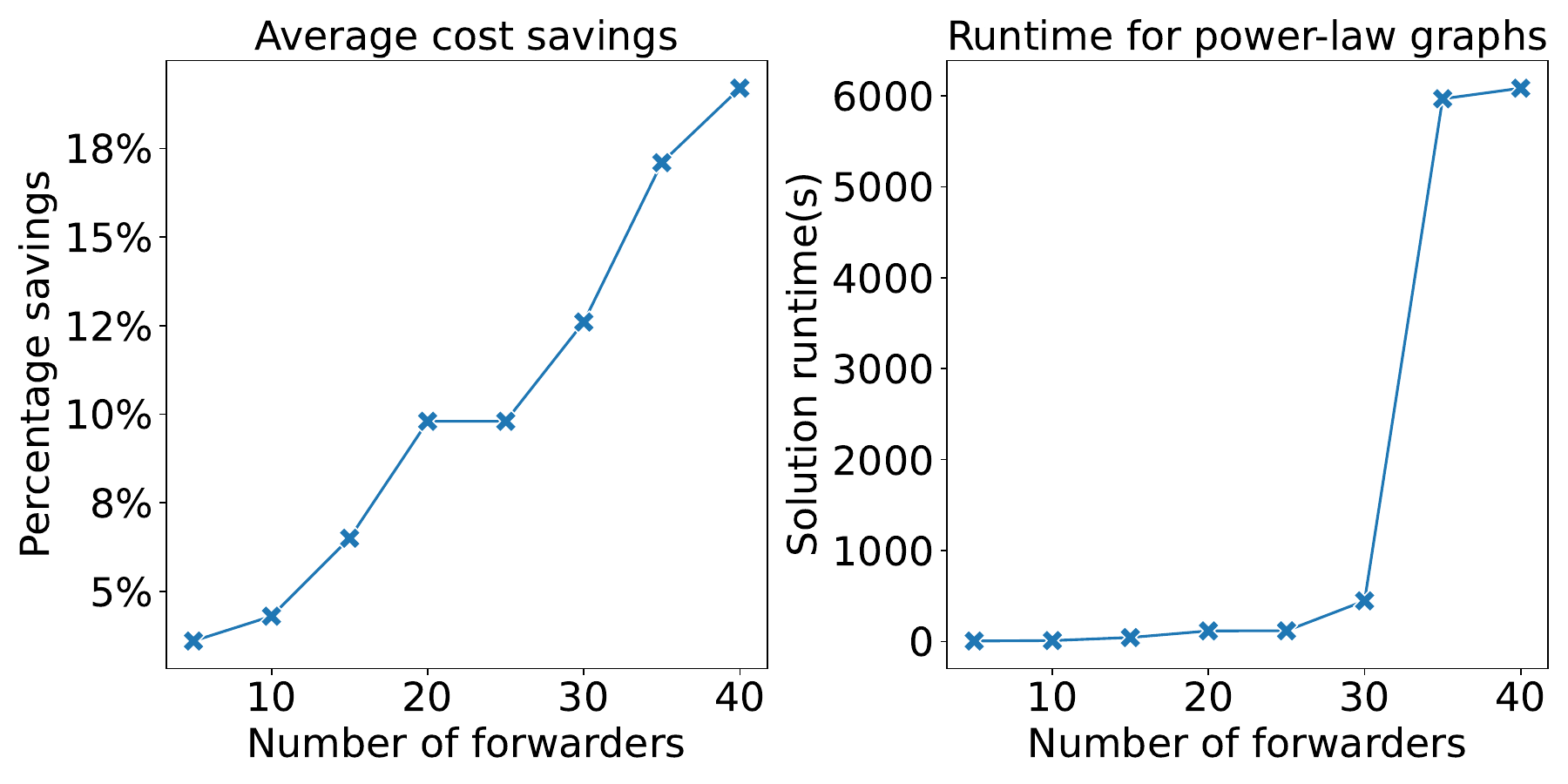}
  \caption{Average savings and runtime for power-law graphs.}
  \label{savings_and_power}
\end{figure}

In the third experiment, we study networks with power-law distribution focusing on cost savings and runtime of FS-LCG. In the first part, we investigate the average cost savings that the forwarders receive when they join the platform as compared to the case when they are on their own. As more forwarders join the platform, they enjoy higher savings, where an average saving of 19\% is observed when 40 forwarders are in the platform (see Figure \ref{savings_and_power}).  In the second part, we investigate the runtime of our approach for the scenario with power-law graphs. While the runtime does increase over the range of the number of forwarders, the Shapley value can still be computed within 1.6 hours for even the largest instance of 40 forwarders (see Figure \ref{savings_and_power}). Here the runtime is still reasonable for practical purposes since such shipment consolidations are usually done as part of short to mid-term planning and they are typically solved every few days rather than on a real-time basis.

\section{Related Work}

Conventionally in cooperative game theory, we assume that each agent could collaborate with any other agent, which then led to the characteristic function being defined on all subsets of agents. However, in many realistic settings, not all agents can cooperate with all other agents. Towards that end, Myerson proposed a cooperative game over a graph called a graph-restricted game. In these games, agents are nodes on the graph and two agents (or nodes) can collaborate only if there is a path between the two nodes. This implies that the characteristic function is only defined for connected coalitions but not for every possible subset of the agents. Several approaches have been proposed to address this issue. \citet{myerson1977graphs} proposed that the payoff of a disconnected coalition is the sum of payoffs of connected coalitions. This leads to the Myerson game. The calculation of Myerson values has also been studied, for example, \citet{bilbao1998values} proposed an explicit formula while \citet{fernandez2002generating} proposed a method based on Harsanyi's dividends. \citet{amer2004connectivity} proposed that the payoff for a disconnected coalition is set to $0$ and that the payoff for a connected coalition is set to $1$. This leads to the connectivity game. \citet{skibski2014algorithms} and \citet{skibski2019enumerating} developed algorithms based on efficient enumeration of induced subgraphs to compute the Shapley value for the Myerson game and the connectivity game. Our work in this paper is similar to these previous works in that we define a game whose characteristic function depends on the underlying graph structure and our main contribution is to exploit this structure to derive an efficient algorithm for computing Shapley value for the game.

In terms of the application domain, our work is most closely aligned with collaborative logistics. The literature on collaborative logistics can be broadly categorized into three main areas: carrier collaboration, shipper collaboration, and forwarder collaboration. Carrier collaboration has been extensively explored, particularly in the context of both Full-Truck-Load (FTL) and Less-than-Truck-Load (LTL) trucking. For instance, \citet{li2015request} introduced a single-lane request approach for FTL trucking, where buyers and sellers submit multiple requests, and a central coordinator selects the most socially beneficial lane for exchange. \citet{lai2017iterative} extended this approach by allowing multiple requests and explored connections to bundle generation and pricing. Freight consolidation is also a prevalent strategy in collaborative transportation. \citet{zhang2018moulin} investigated fair allocation for shippers utilizing consolidation centers, while \citet{lai2022cost} delved into a shipper consortium problem, where logistics service providers optimize routes, consolidate shipments, and allocate costs back to shippers. Collaboration among air carriers and ocean liners is often studied in the context of alliance formation, typically employing cost allocation approaches \cite{agarwal2010network,houghtalen2011designing}. In contrast, research on forwarder collaboration is relatively limited. 

\section{Conclusion}
  
In this paper, we proposed a new class of graph-restricted games called Locally Collaborative Games (LCGs) where the characteristic function has the following property: the marginal contribution of an agent to any coalition only depends on its neighbors. By exploiting this structure, we develop an algorithm called Fast Shapley for Locally Collaborative Games (FS-LCG) for computing Shapley value for LCG. The main idea is that for a given agent, we only need to focus on the induced subgraphs generated by its neighbors to obtain its marginal contribution and count the relevant permutations that result in the given marginal contribution. Our next contribution is to propose the Freight Forwarders Collaboration Game (FFCG) which models how forwarders could collaborate to improve their collective cost efficiency. Since FFCG is an instance of LCG, we apply FS-LCG to compute equitable cost allocation for forwarders participating in the collaboration. For numerical experiments, we investigate three different scenarios. In the first scenario, we note that FS-LCG is faster than the baseline method by the range of 82\% to 93\% under a general setting. Our second experiment involves generating small-word graphs to simulate fragmented markets. The results show that FS-LCG outperforms the baseline with increasing intensity in the number of forwarders or the average degrees. Finally, our third experiment involves generating graphs with power-law degrees to reflect the dominance of a few major market players in the forwarding industry. Under this scenario, forwarders enjoy higher savings when there are more forwarders in the platform, up to a maximum of 18\% savings. Finally, FS-LCG runtime remains competitive and practical even for larger instances.

\section*{Acknowledgments}

This research is supported by the Singapore Ministry of Education (MOE) Academic Research Fund (AcRF) Tier 1 Grant (Project Number: 23-SIS-SMU-017).

\clearpage


\begin{appendices}

\renewcommand\thefigure{A.\arabic{figure}}
\renewcommand\thetable{A.\arabic{table}}
\renewcommand\thealgorithm{A.\arabic{algorithm}}
\renewcommand\theequation{A.\arabic{equation}}

\setcounter{figure}{0} 
\setcounter{table}{0} 
\setcounter{algorithm}{0} 
\setcounter{equation}{0} 

\section{ } 
\section*{Solving the Assignment Problem Efficiently}

In Section 6 of our paper, we mention the Freight Forwarders Collaboration Problem (FFCP) can be solved efficiently by exploiting the decomposability property of the formulation and a greedy approach \cite{10371994}. In this appendix, we describe in detail how this can be done.

\begin{figure}[htb]
  \centering
  \includegraphics[width=3.375in]{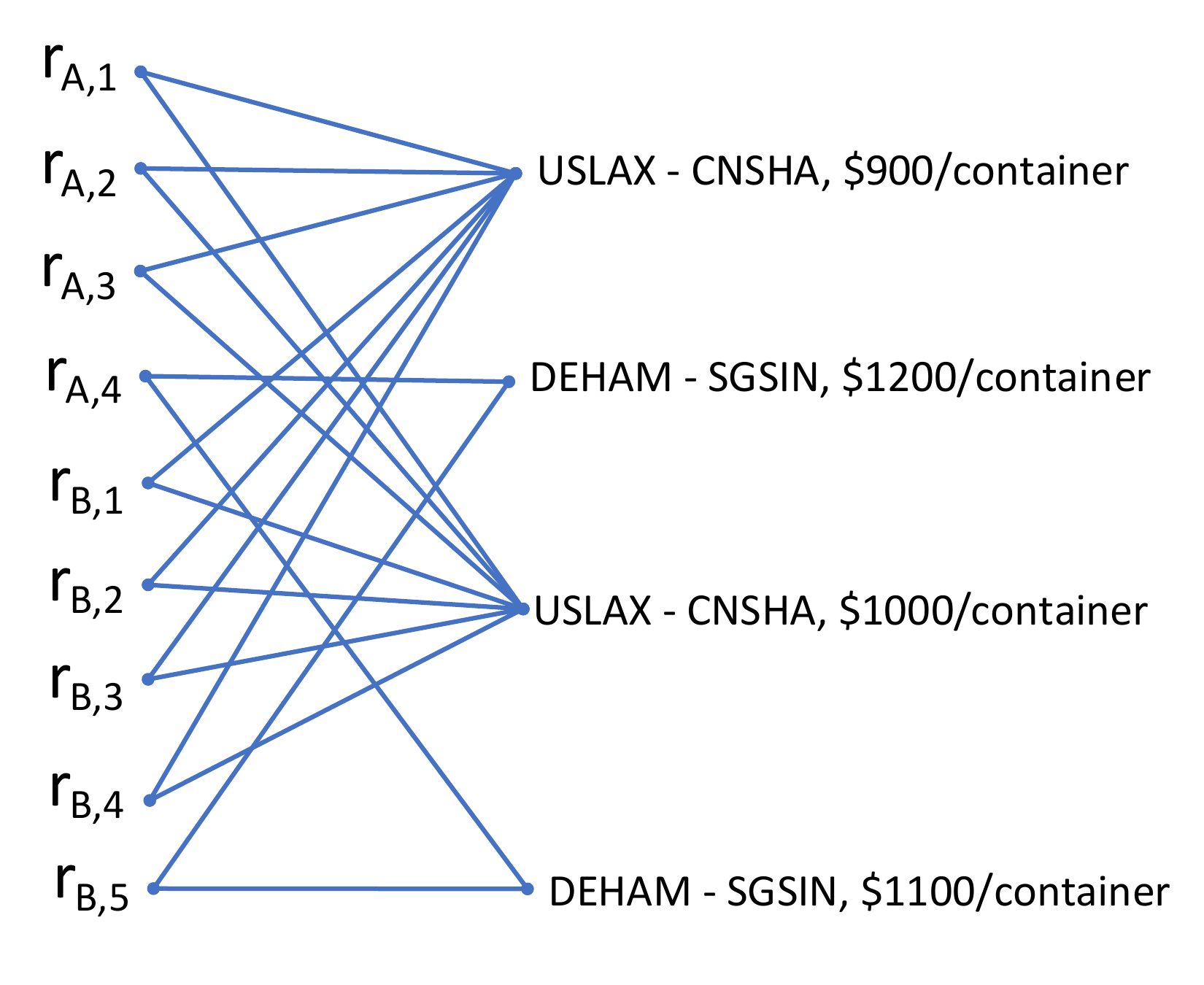}
  \caption{Assigning requests to services in FFCP.}
  \label{graph1A}
\end{figure}
\begin{figure}[htb]
  \centering
  \includegraphics[width=3.375in]{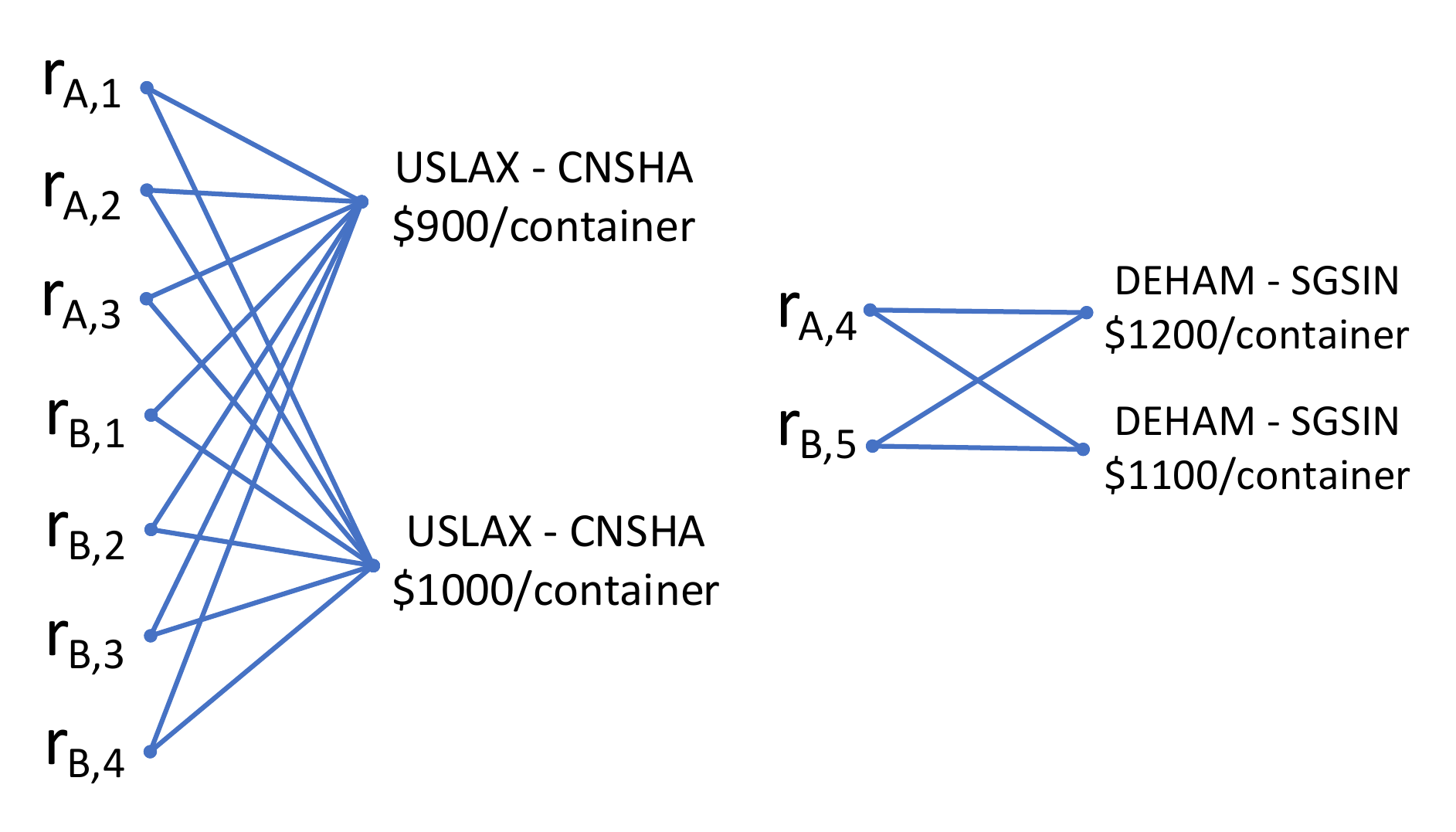}
  \caption{Request-service assignments decomposed into non-overlapping sub-problems.}
  \label{graph2A}
\end{figure}

Note that the request-box assignments can be represented as a graph (see Figure \ref{graph1A}), where vertices on the left represent requests indexed by a forwarder and a request number, and vertices on the right represent a box for a service. Our next observation is that FFCP($R,S$) can be decomposed into non-overlapping sub-problems where each sub-problem deals with a group of services belonging to the same port pairs. This makes sense because if a request is supposed to be shipped on say service USLAX-CNSHA of a forwarder, then the request can be shipped on any other service offered by other forwarders as long as it is serving USLAX-CNSHA. Figure \ref{graph2A} illustrates our point. Instead of solving FFCP($R,S$) in its entirety, we break down the problem into groups of services, each group having services for the same port pair, and solve each sub-problem separately. The total cost is simply the sum of the minimal cost for each sub-problem.


Finally, we turn our focus to solving the sub-problem, which is a special case of FFCP($R,S$) where each request can be assigned to any of the services. We take a two-step approach to solve this special case of FFCP($R,S$). In the first step, we obtain an initial solution by greedily assigning requests to a container with the first fit decreasing heuristic. This gives us a feasible solution and an upper bound on the number of containers needed for each service, $n'_s$. In the second step, we solve the special case of FFCP($R,S$) but the number of containers available for each service is now updated to $n'_s$, as obtained in the first step, instead of the original $n_s$.

In the second step, we solve the following integer linear program exactly.
\begin{align}
    & \min \sum_{s \in S} \sum_{i=1}^{n_s} c_s y^i_s\\
\text{s.t.} & \\
    & \sum_{s \in S} \sum_{i=1}^{n_s} x^i_{r,s} = 1, \forall r \in R, \\
    & \sum_{r \in R}  v_r x^i_{r,s} \le v^{max} y^i_s, \forall s \in S, i=1,...,n_s,\\
    & x^i_{r,s}, y^i_s \in \{0,1\},
\end{align}
where: $R = \{1,2,3,4,5,6,7\}$, $S = \{1,2\}$, $n_1 = 2, n_2=1$, $c_1 = 900, c_2=1000$, $v_1=14, v_2=12, v_3=10, v_4=v_5=v_6=v_7=6$, $v^{max}=30$.

The optimal solution is obtained by assigning the 14 cbm, 10 cbm, and 6 cbm requests to one \$900 container and the remaining requests to another \$900 container. Note that we now use one less container as compared to the greedy approach in the first step.

\end{appendices}

\end{document}